\documentclass[a4paper,10pt,twocolumn]{aastex62}

\hypersetup{linkcolor=red,citecolor=blue,filecolor=cyan,urlcolor=magenta}
\usepackage{natbib}
\usepackage{amsthm}
\usepackage{amssymb}
\usepackage{amsmath}
\usepackage{bm}
\usepackage[utf8]{inputenc}
\usepackage{graphicx}
\usepackage[caption=false]{subfig}
\usepackage{soul}
\usepackage{dblfloatfix}    

\usepackage{lineno}
\usepackage{comment}

\shorttitle{PIC simulation of The Electron-deficit Whistler Instability in the Solar Wind}
\shortauthors{Micera et al.}

\begin{document}

\title{Quasi-parallel anti-sunward propagating whistler waves associated to the electron-deficit in the near-Sun solar wind: Particle-in-Cell simulation}

\correspondingauthor{Alfredo Micera}
\email{alfredo.micera@rub.de}

\author[0000-0001-9293-174X]{Alfredo Micera}
\affiliation{Institut f\"ur Theoretische Physik, Ruhr-Universit\"at Bochum, Bochum, Germany}

\author[0000-0002-0497-1096]{Daniel Verscharen}
\affiliation{Mullard Space Science Laboratory, University College London, Dorking, UK}

\author[0000-0002-2576-0992]{Jesse T. Coburn}
\affiliation{Mullard Space Science Laboratory, University College London, Dorking, UK}
\affiliation{Space Science Institute, Boulder, CO 80301, USA}

\author[0000-0002-5782-0013]{Maria Elena Innocenti}
\affiliation{Institut f\"ur Theoretische Physik, Ruhr-Universit\"at Bochum, Bochum, Germany}

\begin{abstract}

In-situ observations of the solar wind have shown that the electron velocity distribution function (VDF) consists of a quasi-Maxwellian core, comprising most of the electron population, and two sparser components: the halo, which are suprathermal and quasi-isotropic electrons, and an escaping beam population, the strahl. Recent Parker Solar Probe (PSP) and Solar Orbiter (SO) observations have added one more ingredient to the known non-thermal features, the deficit—a depletion in the sunward region of the VDF, already predicted by exospheric models but never so extensively observed.
By employing Particle-in-Cell simulations, we study electron VDFs that reproduce those typically observed in the inner heliosphere and investigate whether the electron deficit may contribute to the onset of kinetic instabilities.
Previous studies and in-situ observations show that strahl electrons drive oblique whistler waves unstable, which in turn scatter them. As a result, suprathermal electrons can occupy regions of phase space where they fulfil resonance conditions with the parallel-propagating whistler wave.
The suprathermal electrons lose kinetic energy, resulting in the generation of unstable waves. The sunward side of the VDF, initially depleted of electrons, is gradually filled, as this wave-particle interaction process, triggered by the depletion itself, takes place.
Our findings are compared and validated against current PSP and SO observations: among others, our study provides a mechanism explaining the presence in the heliosphere of regularly observed parallel anti-sunward whistler waves; suggests why these waves are frequently observed in concomitant with distributions presenting an electron deficit; describes a non-collisional heat flux regulating process.

\end{abstract}

\keywords{Plasma Astrophysics --- Solar wind --- Space plasmas}

\section{Introduction}\label{sec.0}
%\begin{itemize}
%\item The subsonic electrons are %separated into three populations:
%Escaping, Ballistic, Trapped
%\item Presence of the cut-off (Electrons with initial energies smaller than the total electric potential.)
%Boldyrev et al. 2020, Halekas 2022 (?), Bercic 2022 and Heliospheric models
%\item PSP and SOLO observations of deficit (cutoff) strahl (escaping) , core (trapped + ballistic) and halo (where the halo comes from ?))
%\item Halekas et al. (2020) No halo close to the SUN
%\item Štverak et al. (2009) halo appears with the heliocentric distance
%\item Micera et al. 2021 : diffusive particle flux under resonant 
%interactions with waves can occupy the cut-off region of the VDF
%Electron - Deficit instability 
%\end{itemize}

The Parker Solar Probe (PSP, \citet{Foxetal2016}) and Solar Orbiter (SO, \citet{Muller_2013}) missions have provided valuable evidence confirming the fundamental role of electrons in coronal and solar wind dynamics.
Electrons are lighter than ions: their thermal velocity at the corona is large enough for many of them to escape the Sun's gravity. An ambipolar electric field is then established, that decelerates electrons and accelerates protons, as to maintain equal electron and ion fluxes in the radial direction~\citep{meyer2007basics}. The existence of this ambipolar field has been predicted and explained in the context of exospheric models~\citep{jockers1970solar, Lemaire_1971, maksimovic2001exospheric, zouganelis2005acceleration}, which assumes collisioness particle dynamics above a reference level called the exobase. It has also been studied in global-scale models, where increased levels of collisionality are used as a proxy for wave/particle interaction \citep{lie1997kinetic, pierrard1999electron, landi2003kinetic}.

The ambipolar electric field deeply influences electron circulation patterns in interplanetary space and hence the shape of the electron VDFs. 
Three electron populations are present in the solar wind: escaping, trapped and ballistic electrons. Escaping electrons have energy larger than the asymptotic ambipolar potential energy, and stream away to increasingly large radial distances. Both ballistic and trapped electrons are turned back towards the Sun by the ambipolar electric potential. Ballistic electron are the ones which "fall back" into the collisional coronal reservoir. Trapped electrons are the one which are again turned back, this time towards increasing radial distance, by the mirror force in the sunward-increasing magnetic field~\citep{Lemaire1973, scudder1996dreicer, Pierrard_1996, Maksimovic1997, meyer1998electron, landi2003kinetic, zouganelis2005acceleration, boldyrev2020electron}.

The typical electron VDF observed in the solar wind, composed of three electron populations (core, strahl and halo~\citep{Feldman1975, Rosenbauer1977, Pilipp1987, Maksimovic2005, Stverak_2009, Halekas2020}), is a direct consequence of this large scale processes. Ballistic and trapped electrons form the core, escaping electrons constitute the strahl.  Scattered strahl electrons give rise to the halo \citep[e.g.][]{Stverak_2009}. Strahl-to-halo scattering has been observed in a number of kinetic models~\citep{Vocks2005, Jeong-Seong2020, tang2020numerical}, including fully kinetic Particle-in-Cell simulations~\citep{Micera2020b, Micera_2021}. It is also indirectly confirmed by the anti-correlation between the fractional density of the strahl and halo populations observed e.g. in \citet{Maksimovic2005, Stverak_2009} at $r>0.3~ au$, where $r$ is the heliocentric distance, and more recently by \citet{Halekas2020} and \citet{Bercic2020} in PSP data. Interestingly, one can find an early, indirect observation of strahl-to-halo scattering in \citet{Scime1994}. There, the radial heat flux evolution in Ulysses data ($1<r<5\; au$) is compared with expectations from collisionless expansion of a suprathermal population (called there ``halo") along magnetic field lines. The observed heat flux exhibits a $\propto r^{-3}$ radial dependence, decreasing more sharply than the sole effect of expansion. This is compatible with a scenario where the heat flux-carrying strahl reduces faster than what expected from expansion alone, as a result of strahl-to-halo scattering processes. A radial evolution of strahl density faster than the so-called ``spiral" expansion (more appropriate for the strahl population than ``radial" expansion, following the terminology in \citet{Stverak_2009}) is directly observed by \citet{Stverak_2009}.

In \citet{Micera_2021}, fully kinetic expanding-box simulations run with the semi-implicit EB-iPic3D code \citep{Innocenti2019, Innocenti2020}  demonstrate halo formation from scattering of strahl electrons due to the oblique whistler heat flux instability. The simulation is initialized with a (stable) electron VDF and plasma parameters that accurately reproduce those measured during PSP Encounter I. 
Radial expansion self-consistently drives the solar wind in a regime where it is unstable with respect to whistler heat flux instability.
\citet{Cattell2021} provide strong observational evidence for this numerically predicted scenario: the direct evidence for pitch angle scattering of strahl electrons by narrowband whistler-mode waves in PSP observations.

The presence of the ambipolar electric potential in interplanetary space leaves a distinct signature in the electron VDFs, namely the so-called electron deficit in the sunward magnetic-field-aligned direction. The "missing" returning electrons are those energetic enough to escape the electrostatic potential. Early deficit observations in Helios data down to 65 $R_s$~\citep{Pilipp1987} are recently corroborated by PSP~\citep{Halekas2020, Halekas_2021_deficit, bervcivc2021ambipolar} and SO observations~\citep{bervcivc2021whistler, Coburn2024}.
In \citet{bervcivc2021ambipolar}, a number of VDF measurements collected between 20.3 $R_s$ and 85.3 $R_s$, with $R_s$ the solar radius, during PSP Encounters 4 to 7 are used to calculate the location in energy of the deficit ("cut-off energy") and, from that, the ambipolar potential between a specific location and its asymptotic value. They measure a radial dependence of $r^{-0.66}$. In~\citet{Halekas_2021_deficit}, the statistical properties of the deficit are highlighted. The deficit is observed more frequently closer to the Sun (below $0.2~ au$ the deficit occurs in $60$ to $80~ \%$ of observations while clear signs of its presence appear less frequently at larger distances), with lower fractional halo density, smaller electron parallel beta, lower collisional age, more anisotropic core distributions. This suggests causality or correlation between the processes that isotropize electron VDFs (e.g. strahl-to-halo electron scattering) and those which erase deficit signatures.  
\citet{Halekas_2021_deficit} suggest two possible mechanisms for deficit erasure by collisionless processes. One possibility is a multi-step process, similar to the one proposed by \citet{Micera2020b, Micera_2021}, which scatters strahl electrons into the halo and then relaxes into parallel whistler waves. These waves would further isotropize the distribution by scattering halo electrons at all pitch angles, thus erasing the deficit. A second possibility is an instability triggered by an unstable deficit, as suggested by~\citet{bervcivc2021whistler} and further explored by~\citet{Coburn2024}. There, quasi-parallel right-hand polarised whistler waves are observed in SO field data at 112 $R_s$, in the presence of the electron deficit. Resonance condition analysis supports the hypothesis of an anti-sunward, quasi-parallel whistler instability driven by electrons scattering from higher to lower energies. Such an instability would contribute to erase the electron deficit.

In \cite{Micera2020b} it has not been ascertained whether the parallel whistler waves that scatter nascent halo electrons at the highest pitch angles are a result of the relaxation of the oblique whistler heat flux instability.
Alternatively, these parallel whistler waves might form independently, triggered by changes in the electron VDF, as suggested in \cite{bervcivc2021whistler}.\\
In addition, in \cite{Micera_2021} most of the parallel whistler waves are sunward-directed, being essentially formed from the relaxation of the whistler heat flux instability triggered by the strahl. However, the waves that resonantly scatter the sunward deficit can only propagate anti-sunward \citep{bervcivc2021whistler}.

In this work, we simulate via fully kinetic Particle-in-Cell simulation a VDF resembling that obtained in \citet{Micera2020, Micera_2021}, to elucidate the relation between strahl-to-halo scattering, quasi-parallel whistler waves and electron deficit erasure. To do so, we initialize our simulation with a configuration that is not unstable to the whistler heat flux instability to isolate instabilities directly driven by the electron deficit, if they at all occur.

This letter is organized as follows: in section \ref{sec.2}, we show the simulation setup and describe in detail our initial conditions, motivating them in the context of solar wind electron dynamics. In section \ref{sec.3}, we show the results of the simulation by analysing waves, particle distributions and their mutual interactions. Finally, in section \ref{sec.5}, we discuss our results, compare and validate them against recent observations conducted by PSP and SO and draw our conclusions.

\section{Setup of the PIC simulation}\label{sec.2}

%\begin{comment}
\begin{figure}[htbp]
    \centering
    \includegraphics[width=\columnwidth]{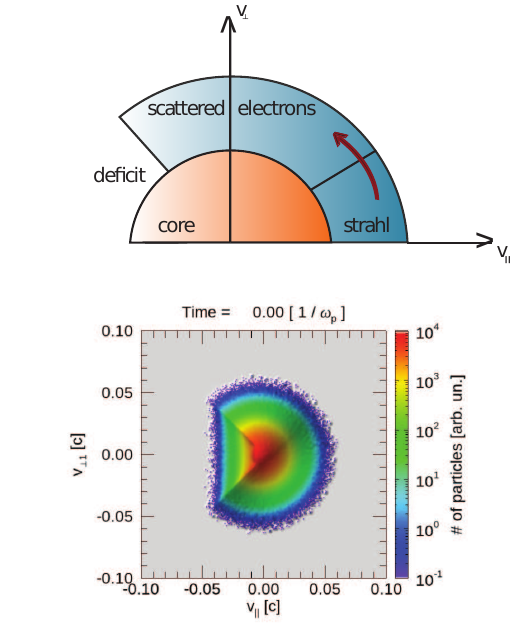}
    \caption{Schematic example of the electron distribution function used to initialize the simulation (a). The orange area denotes the region of the phase space that can be occupied by trapped and ballistic electrons, while the blue area by escaping and scattered electrons. Electron VDF $f_e = f(v_{\parallel}, v_{\bot_{1}})$ at $t_0=0$ (b). The phase space is integrated over $v_{\bot_{2}}$.}
    \label{fig_1}
\end{figure}
%\end{comment}

We perform a $2D$ PIC simulation with iPic3D \citep{Markidis2010, Innocenti2017b}, a fully-kinetic code that uses a semi-implicit scheme \citep{Lapenta2006} to couple the Maxwell's equations governing electromagnetic fields and the equations of motion that describe the dynamics of particles.
Thanks to the semi-implicit approach, small temporal and spatial scales can be retained, without the requirement to resolve the Debye length and the inverse electron plasma frequency as is the case in explicit codes \citep{Hockney} and thus including electronic scales and at the same time employing a domain of the order of several ion skin depth.

We model a non-collisional plasma, consisting of electrons and protons, embedded in an initially uniform magnetic field. The initial background magnetic field is of magnitude $0.00045~ G$ and is directed along the $x$-direction $(\mathbf{B_0} =B_0~ {\hat{e}_x})$, which is  the direction we denote as parallel ($\parallel$). Positive $x$ is the anti-sunward direction. $y$ and $z$ are the two perpendicular directions that we denote as  $\perp_1$ and $\perp_2$, respectively.\\

The units used for normalisation are as follows:
for velocities, the speed of light $c$, for lengths the proton skin depth $d_p = c / \omega_{p}$ and for the time the inverse of the proton plasma frequency $\omega_{p} = \sqrt{4 \pi e^2 n_p / m_p}$, with $n_p$ the proton number density, $m_p$ the proton mass and $e$ the elementary charge.

The computational domain consists of a $2D$ box of size $L_{x, y} = 8~ d_p$, discretized with $512^2$ grid points so that the spatial resolution is $\Delta_{x, y} = d_p / 64$. The time step used is $0.05~ \omega_{p}^{-1}$. We use a realistic proton-to-electron  mass ratio ($\mu = m_p/m_e = 1836$) and $1024$ protons and $4096$ electrons per cell. 
The choice of such a set-up results from convergence tests conducted to find a resolution and number of particles such that the energy is conserved almost perfectly (see Figure \ref{fig_2}) and to have satisfactory statistics for the region of phase space occupied by suprathermal electrons.

The initial electron distribution consists of an electron core (in orange in the sketch image in Figure \ref{fig_1} (a)), plus suprathermal electrons (strahl + nascent halo, depicted in blue). 
Here, the suprathermal electrons are the sum of escaping electrons (the strahl) plus those that are scattered during the interaction between the strahl and oblique whistler waves and thus acquire higher perpendicular velocities \citep{Micera2020b, Cattell2021, Micera_2021}.  
The scattered electrons can populate the suprathermal trajectories around and up to over 90 degrees pitch angle and can be seen as incipient halo electrons.\\
Underlying our initial condition is the process whereby in a deficit-strahl system, due to the generation of oblique whistler waves, electrons with high $v_{\perp}$ and $v_{\parallel} \leq 0$ can be obtained. This process has already been simulated in the framework of the solar wind in \cite{Micera2020b, Micera_2021} and in other applications in which the electron distribution function is modelled through the employment of PIC simulations \citep[e.g.][]{Komarov2018,Roberg-Clark2019}.
Here we start from the scenario where part of the strahl electrons have already been scattered, and the halo is not fully formed, and thus the sunward deficit is not fully filled by the suprathermal electrons. 
This is a configuration that can be commonly observed at certain heliocentric distances where the three typical supratheraml features of the electron VDF can coexist: the strahl, a tenuous halo and the deficit \cite[e.g.][]{Halekas_2021_deficit, bervcivc2021ambipolar}.

The choice of employing such an electron distribution function, which is the product of a transient regime, to initialize our simulations derives from the interest to understand the interplay between the various suprathermal components of the electron VDF: in particular, whether the wave-particle interactions that lead to the formation of the halo at the expense of the strahl also have an effect in filling the deficit as the solar wind travels through interplanetary space.

For these reasons, the initial electron VDF consists of a drifting Maxwellian, from which we cut a solid angle in the sunward region of the phase space: the "missing" electrons model the electron deficit. The drifting Maxwellian is defined as follows:

\begin{equation}
     \begin{aligned}
f_e (v_{\parallel}, v_{\perp}, t=0) = \frac{(2 \pi)^{- 3/2}}{w^2_{\perp e} w_{\parallel e}} \exp \left(-\frac{v_{\perp}^2}{2 w_{\perp e}^2} - \frac{(v_{\parallel} - u_e)^2}{2 w_{\parallel e}^2} \right), 
    \label{1}   
     \end{aligned}     
\end{equation}
with $w_e = \sqrt{k_B T_e / m_e}$ the electron thermal velocity, $k_{B}$ the Boltzmann constant and $T_e$ the electron temperature.
The deficit in the electron VDF (of which we observe a $2D$ projection in Figure \ref{fig_1} (b) and a $3D$ view in Figure \ref{fig_3} (a)), is modelled by excluding from the distribution the electrons whose velocities satisfy the following quadratic law:
\begin{equation}
     \begin{aligned}
v_{\parallel} < -  p~ \sqrt{ (v_{\perp 1}^2 + v_{\perp 2}^2)}.
    \label{2}   
     \end{aligned}     
\end{equation}
The parameter $p$ represents the free parameter through which we decide the angle at which to cut our VDF in the plane. In this letter, we have chosen to use $p=1$ to obtain thus a cut between the angles $\alpha = 90^{\circ} + \mathrm{arctan} (\sqrt{(v_{\perp 1}^2 + v_{\perp 2}^2}) / v_{\parallel}) = 144^{\circ} $ and $\alpha_1 = 360^{\circ} - \alpha = 216^{\circ}$.

Protons are assumed to be initially isotropic and Maxwellian. The proton and electron temperatures are chosen so that $\beta_p = 1.7$ and $\beta_{e \parallel} = 1.5$, with $\beta_{j \parallel} = 8 \pi n_{j} k_B T_{j \parallel} /B_0^2$ and subscript $j$ denotes the species ($p, e$).
We assume that their drift velocity is zero ($u_i = 0$).
To ensure that the zero net-current condition is satisfied, we impose a sunward drift on the electron distribution ($u_e = -0.004~c$), which balances the current due to the initial VDF choice. This adjustment is crucial to maintain zero net current. Indeed the absence of electrons associated with the electron deficit in the sunward direction would result in an antisunward-directed drift speed if the peak of the Maxwellian were centered at $v_{\parallel}/c=0$.
To ensure that the net plasma current at initialisation is zero, we added a negative drift to the electron distribution so that the total electron current at initialization is zero.
Additionally, the plasma is required to satisfy the quasi-neutrality condition ($n_{p} = n_{e}$), and we ensure that the proton and electron densities are equal.

\section{{Simulation Results}}\label{sec.3}
\begin{figure}[htbp]
    \centering
    \includegraphics[width=\columnwidth]{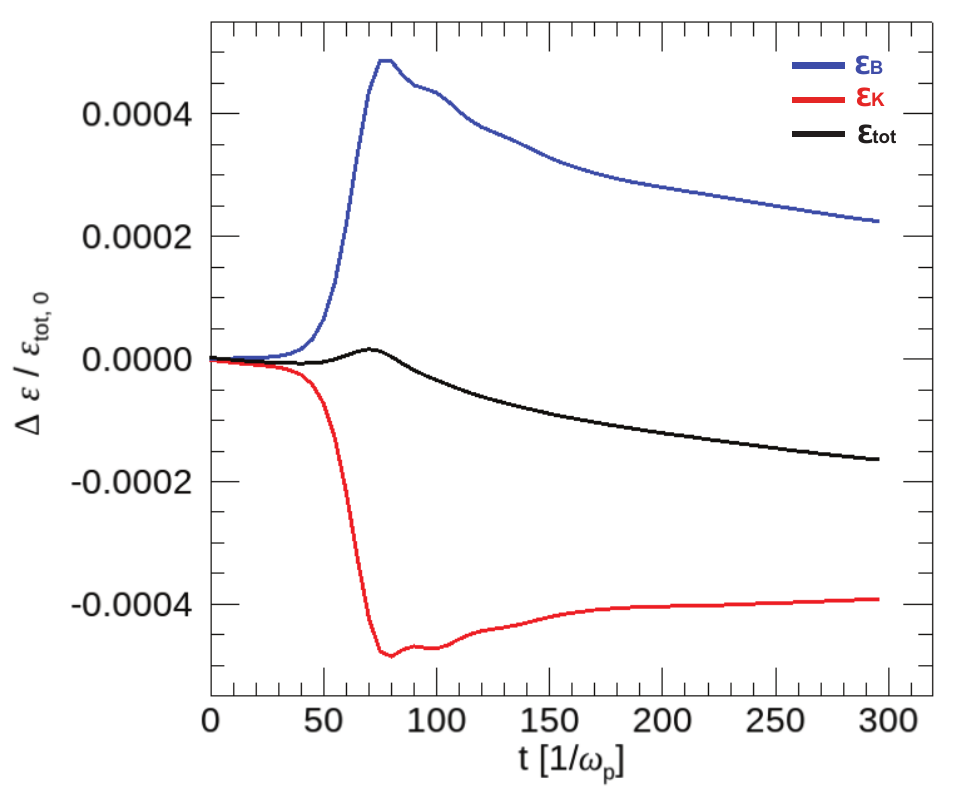}
    \caption{Temporal evolution of the normalised variation of magnetic energy (blue), kinetic energy (red) and total energy (black).}
    \label{fig_2}
\end{figure}

We let the plasma evolve from its initial condition, described in section \ref{sec.2}, and measure the energy exchanges within the simulated system. Figure \ref{fig_2} shows the temporal evolution of the variation of magnetic energy, kinetic energy and total energy (electromagnetic energy plus kinetic energy) in the simulation. 
All energy variations shown in Figure \ref{fig_2} are offset by their initial values and normalised with respect to the total energy of the system at the initialisation:
\begin{equation}
     \begin{aligned}
\Delta \epsilon / \epsilon_{tot, 0} =  ( \epsilon - \epsilon_{0} ) / \epsilon_{tot, 0}.
    \label{3}   
     \end{aligned}     
\end{equation}

The blue curve represents the evolution over time of the normalised magnetic energy variation, with magnetic energy evaluated as:
\begin{equation}
     \begin{aligned}
     \epsilon_B  = \frac{1}{8\pi} \int_V (B_x^2 + B_y^2 + B_z^2)~  \, dV.
     \label{4}   
     \end{aligned}     
\end{equation}

The magnetic energy, after an initial phase in which it remains constant, presents an exponential growth starting at about $t = 50~ \omega_p^{-1}$.
The magnetic energy reaches a peak at time $72~ \omega_p^{-1}$, after which it smoothly decreases until the end of the simulation.\\
The curve in red, i.e. the normalised variation of the kinetic energy of the plasma ($\Delta \epsilon_K / \epsilon_{tot, 0} $), with kinetic energy evaluated as:
\begin{equation}
     \begin{aligned}
\epsilon_K  = \frac{1}{2} m_e \vec{v}_e \cdot \vec{v}_e + \frac{1}{2} m_p \vec{v}_i \cdot \vec{v}_i,
\end{aligned}     
\end{equation}
%one can see that the energy gained by the magnetic field is lost by the particles
is antisymmetric to the curve representing the magnetic energy variation over time. This is due to an interchange of kinetic and magnetic energies in the simulated system (the energy gained by the magnetic field is lost by the particles and vice versa). After an initial quasi-stationary phase, an electromagnetic instability is triggered, that leads to a process of wave amplification at the expense of the kinetic energy of the electrons. Due to their initial configuration, the electrons have free energy that is gradually transferred to the electromagnetic fields during the growth phase of the instability.  
Once the magnetic energy has peaked, we enter the saturation phase of the instability, where the magnetic energy is converted back into kinetic energy. In Figure \ref{fig_2}, the black curve represents the relative change of the total energy with respect to its value at initialisation ($\Delta \epsilon_{tot} / \epsilon_{tot, 0}$) with:
\begin{equation}
    \begin{aligned}
\epsilon_{tot} = \epsilon_{B} + \epsilon_{K} + \epsilon_{E}, 
\end{aligned}     
\end{equation} 
and 
\begin{equation}
    \begin{aligned}
\epsilon_E  = \frac{1}{8\pi} \int_V (E_x^2 + E_y^2 + E_z^2)~  \, dV,
\end{aligned}     
\end{equation} the plasma electric field energy.\\
The total energy remains nearly constant throughout the duration of the simulation. 
The small amount of numerical cooling we observe is a characteristic of non energy-conserving semi-implicit discretizations, which tend to remove energy from the system, while explicit discretizations tend to introduce numerical heating. An energy-conserving semi-implicit discretization has recently been introduced in \cite{Lapenta2017}, but has not been used in the present work.\\
We remark that the physical significance of the simulation is ensured by the fact that the amount of energy converted at the end of linear phase ($\approx 0.05~\%$ at the peak) is well above the amount of numerical cooling both the at the same time ($\approx 0.005~\%$ at $t = 72~ \omega_p^{-1}$) and also at the end of the simulation ($\approx 0.015~\%$ at $t = 300~ \omega_p^{-1}$).

%\begin{comment}
%\end{comment}

\subsection{Temporal evolution of the electron VDF}

\begin{figure}[htbp]
    \centering
    \includegraphics[width=\columnwidth]{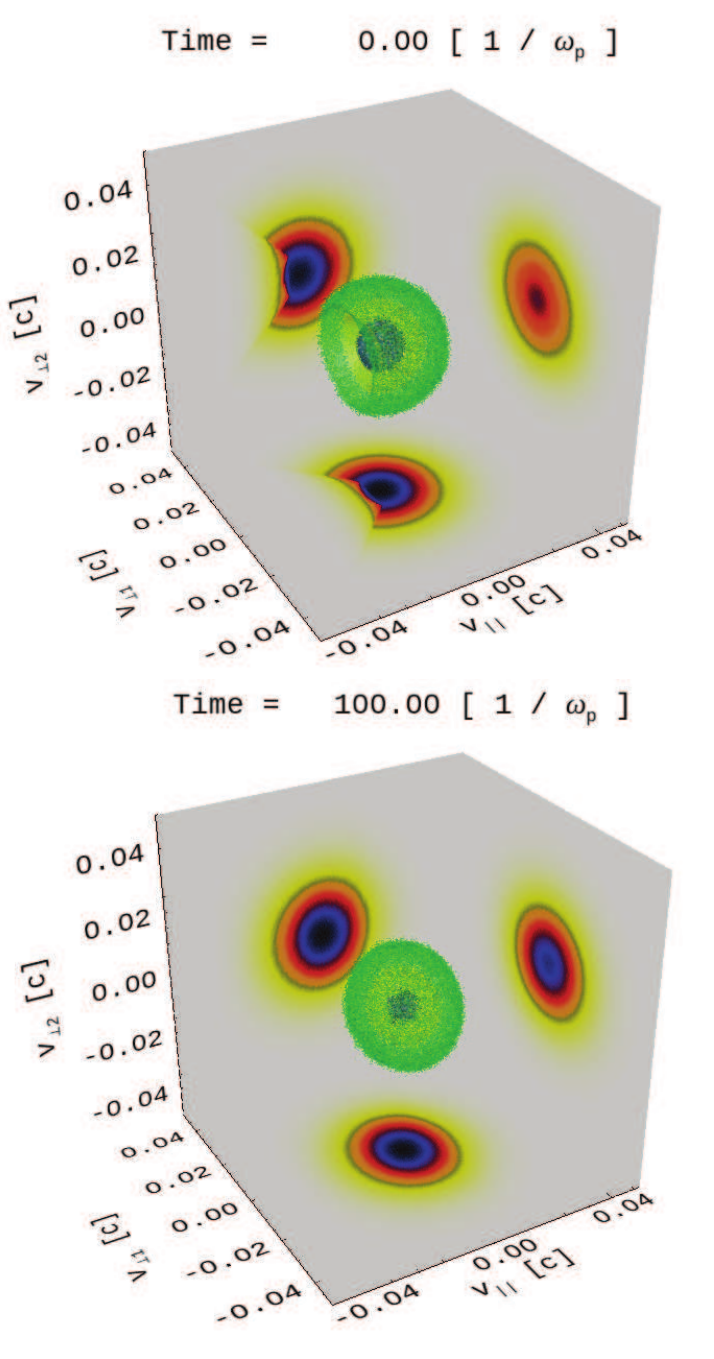}
    \caption{Electron VDF $f_e = f(v_{\parallel}, v_{\bot 1}, v_{\bot 2} )$ at $t_0=0$ (a) and $t = 100~ \omega_{p}^{-1}$ (b). This figure is complemented by Animated Figure 1, which dynamically illustrates the evolution of the electron VDF throughout the simulation.}
    \label{fig_3}
\end{figure}

To study how electromagnetic fluctuations affect the plasma, in Figure \ref{fig_3} we plot the electron distribution function $f_e (v_{\parallel}, v_{\perp 1}, v_{\perp 2} $) at the beginning of the simulation and at the end of the linear growth phase of the instability. 
Figure \ref{fig_3} (a) shows a $3D$ view of the initial electron VDF shown in Figure \ref{fig_1} (b) in a $2D$ plane. We see the presence of a dense, isotropic core and the high-energy features characteristic of electron VDFs in the solar wind: the dynamic deficit-strahl-halo system.\\
In Figure \ref{fig_3} (b), we show how the electron VDF is modified after the generation of the instability.  In particular, the interaction between the electrons and the generated waves results in the filling of the electron deficit and to an electron distribution that is quasi-isotropic during the saturation phase of the instability. \\
Animated Figure 1: the animation illustrates the temporal evolution of the electron VDF over the duration of the simulation. It starts at the beginning of the simulation, progresses through the onset of wave generation, and continues to the saturation phase. The interaction with the generated waves is shown to fill the electron deficit and isotropize the distribution. The animation lasts 5.42 seconds and is available online as supplementary material.\\
An electron distribution with a deficit, in the presence of suprathermal electrons (some of them with a negative parallel velocity component as a consequence of previous scattering) leads to an electromagnetic instability. The fluctuations generated by the instability result in wave-particle interactions such that the deficit is gradually filled.
Protons are not shown because they are initialized in an equilibrium situation and remain largely unperturbed throughout the duration of the simulation.

%\begin{comment}
%\end{comment}

\begin{figure}[htbp]
    \centering
    \includegraphics[width=\columnwidth]{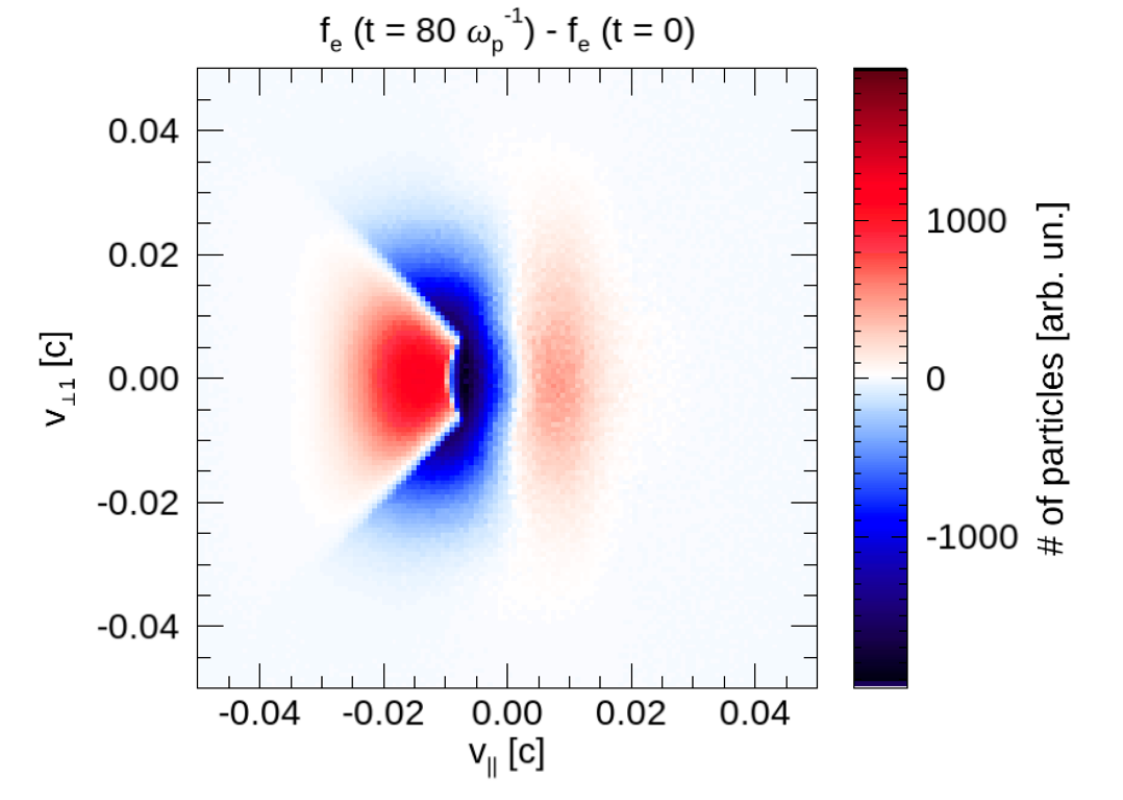}
    \caption{Difference between the electron velocity distribution function near the peak of the instability and that at time zero ($f_e (t = 80 \omega_{p}^{-1}) - f_e (t = 0)$).}
    \label{fig_4}
\end{figure}

We show in Figure \ref{fig_4} the difference between the electron VDF near the peak of the instability ($f_e (t = 80~ \omega_p^{-1})$) and at time zero, to understand which electrons are scattered into the vacant sunward region. The electrons affected by scattering processes with the waves generated during the instability occupy a specific region of phase space: electrons adjacent to the deficit, due to wave-particle resonance interactions, are scattered into a region of phase space that was not initially populated by electrons.

%\begin{comment}
%\end{comment}

\subsection{Nature of the electromagnetic waves}

\begin{figure}[htbp]
    \centering
    \includegraphics[width=\columnwidth]{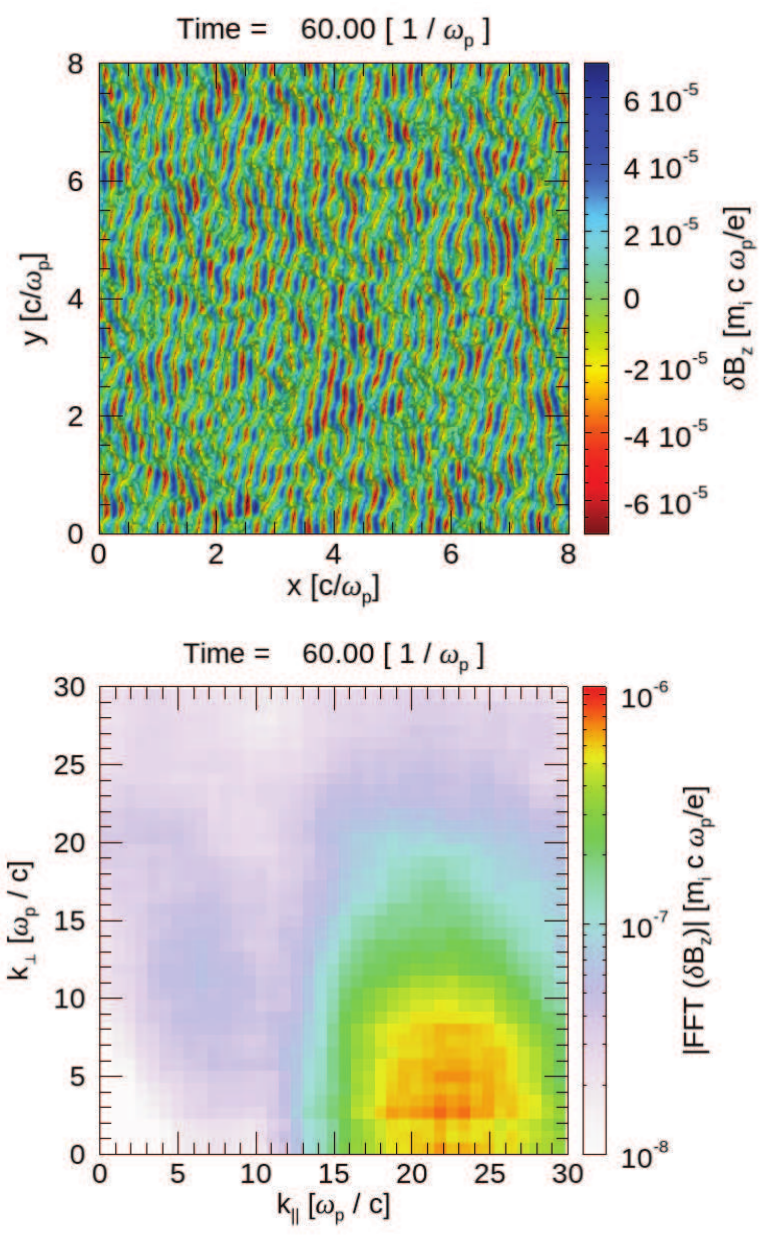}
    \caption{Out-of-plane magnetic field fluctuations during the instability growing phase ($\delta B_z (t = 60~ \omega_{p}^{-1})$) (a) and its FFT (b). This figure is complemented by Animated Figure 2, which dynamically shows the temporal evolution of $\delta B_z (t)$ in the $x - y$ plane during the simulation.}
    \label{fig_5}
\end{figure}

We now investigate in detail the nature of the instability we observe and the electromagnetic waves it produces. 
Figure \ref{fig_5} (a) shows the transverse magnetic fluctuations in the $x - y$ plane during the growing phase of the instability at $t = 60~ \omega_{p}^{-1}$. The fluctuations are calculated as: $\delta B_z (t) = B_z (t) - B_0 $.
The waves have a direction of propagation that is mainly parallel to $\mathbf{B_0} = B_0~ {\hat{e}_x}$, and the box is large enough to contain multiple  oscillations of the fastest-growing waves. 
Animated Figure 2: The animation illustrates the temporal evolution of $\delta B_z (t)$ in the $x - y$ plane throughout the simulation. It begins at $t = 0$, progresses through the growing phase of the instability, and shows the development and propagation of wave crests moving from left to right of the spatial domain. The animation lasts 4.40 seconds and is available online as supplementary material.\\
Figure \ref{fig_5} (b) shows the power spectrum of the out-of-plane magnetic field fluctuations, obtained with a Fast Fourier Transform $FFT (\delta B_z (t))$, in the $k_{\parallel} - k_{\perp}$ plane, where $k_{\parallel}$ and $k_{\perp}$ are the wave vectors parallel and perpendicular to $\mathbf{B_0}$, with $k = 2 \pi / \lambda$ with $\lambda$ the wavelength. 
The instability leads to the generation of unstable electromagnetic waves propagating in a quasi-parallel direction to the background magnetic field. In particular, the fastest growing modes are concentrated between $20~ d_p^{-1}< k_{\parallel} < 26~ d_p^{-1}$ and $0 <k_{\perp} < 6~ d_p^{-1}$, with an angle of propagation of most unstable waves ranging from zero up to $20^\circ$ degrees with respect to the background magnetic field. 
The simulation box we have chosen is capable of containing more than $25$ oscillations of the fastest growing mode, considering it to be characterised by a wavelength $\lambda = 2 \pi / (20~ {d_p^{-1}})$.
The waves migrate to progressively lower $k_{\parallel}$ as the deficit is filled (not shown in this letter).

%\begin{comment}
%\end{comment}

\begin{figure}[htbp]
    \centering
    \includegraphics[width=\columnwidth]{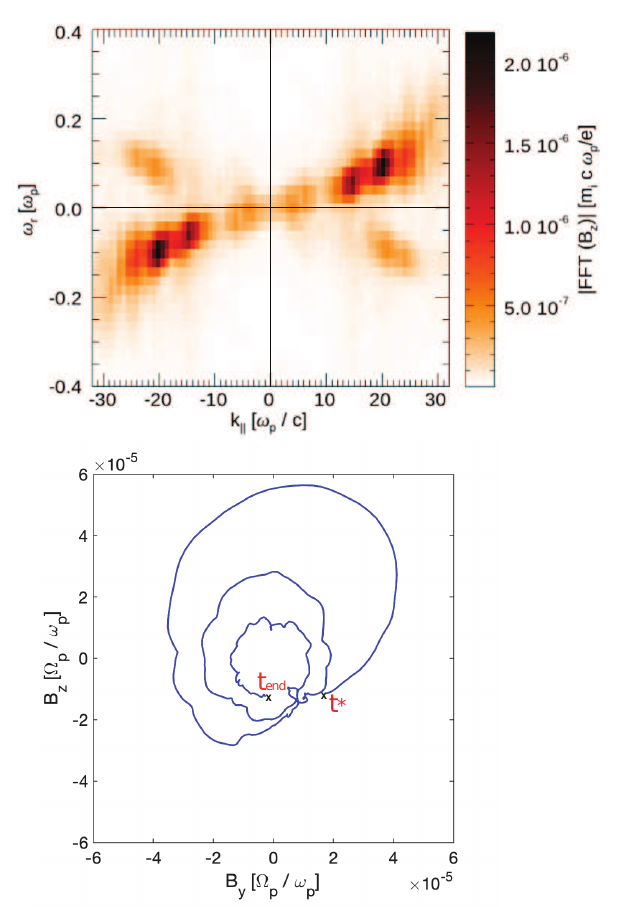}
    \caption{Spacetime Fourier power spectrum $k_{\parallel}-\omega_r$ of predominately anti-sunward whistler waves propagating along the background magnetic field direction (a). Hodogram of right-hand polarised whistler waves (b). $B_x$ is directed out of the page.}
    \label{fig_6}
\end{figure}

Figure \ref{fig_6} (a) shows the spacetime Fourier power spectrum, i.e. the frequency with which the waves propagate as a function of the wave vector $k_{\parallel}$.
We first note that most waves propagate away from the Sun, i.e. in the positive, anti-sunward direction.
This is in line with what has been described above regarding the time evolution of the electron VDF: the electrons that resonate with the waves generated during the instability are those that move in the opposite direction to the waves, and are scattered and go to fill the sunward deficit.
From Figure \ref{fig_6} (a), we observe that the waves propagate at a frequency between $0.025~ \omega_{p}^{-1}$ and $0.13~ \omega_{p}^{-1}$. Our electron gyrofrequency ($\Omega_{e} = e B_0 / (m_e c)$) normalized to the proton (electron) plasma frequency is $\Omega_{e}/ \omega_{p} = 0.42$ ($\Omega_{e}/ \omega_{e} = 0.01$), hence the range of frequencies at which most waves propagate fulfils: $0.06~ \Omega_{e} \leq \omega_r \leq 0.3~ \Omega_{e}$.
The frequencies at which the waves propagate are characteristic of fast-magnetosonic/whistler waves ($\omega_r < \Omega_{e}$) \cite[e.g.][]{Stansby2016}.
The cyclotron resonance condition for parallel-propagating whistler waves \citep{Verscharen_2019} is given by: 
\begin{equation}
     \begin{aligned}
\omega_r - \Omega_{e} = k_{\parallel}~v_{\parallel}. \label{eq.6}  
     \end{aligned}     
\end{equation}
Since $\omega_r < \Omega_{e} $ for whistler waves, Eq.~\ref{eq.6} demands that the resonance interaction only occurs when $k_{\parallel}~v_{\parallel} < 0$. 
This means that parallel whistler waves only scatter electrons that travel in the opposite direction to the waves, in this case, we have positively (anti-sunward) propagating waves which scatter electrons with negative (sunward) parallel velocity.

Figure \ref{fig_6} (b) depicts the wave hodogram, obtained by plotting $B_y$ vs $B_z$ at the centre of the domain $x = y = 4~d_p$ as a function of time, starting from the end of the quasi-stationary phase ($t^{*} = 50~ \omega_p^{-1} $) to the end of the simulation ($t_{end} = 300~ \omega_p^{-1}  $).
We observe that the wave is almost circularly and purely right-hand polarised (the $x$-axis points out of the page), which is again consistent with our interpretation of these waves as parallel-propagating whistler waves.

We therefore conclude that a distribution function that is in line with those commonly observed in the heliosphere near the Sun, where an electron deficit is commonly present, leads to the generation of quasi-parallel, right-hand circularly polarised waves which propagate away from the Sun with frequencies $\omega_r < \Omega_{e}$.
We identify these waves as anti-sunward propagating whistler waves.

%\begin{comment}
%\end{comment}

\section{Discussion and Conclusions}\label{sec.5}
We explore via a fully kinetic simulation a possible scenario for the erasure of the sunward electron deficit. While already predicted by exospheric models \citep[e.g.][]{Lemaire1973, Maksimovic1997}, systematic observations of the electron deficit have been made possible only by the recent Parker Solar Probe and Solar Orbiter campaigns \citep[e.g.][]{Halekas_2021_deficit, bervcivc2021whistler}. 

In our simulation, unstable quasi-parallel anti-sunward whistler waves are generated by the departure from the thermal equilibrium of the electron VDF due to the presence of the deficit itself. 
The waves we observe, as characteristic for whistler waves, have a frequency $\omega_r < \Omega_{e}$ and since they propagate along the direction of the magnetic field they exhibit right-hand circular polarisation.
We show that the simulated whistler instability resonantly scatters electrons from neighbouring regions in phase space into the deficit, effectively erasing it. 

This work proposes a possible mechanism that leads to the generation of quasi-parallel whistler waves in the solar wind. These waves propagate away from the Sun with a frequency of the order of $0.1 \approx \Omega_{e}$. 
This is of high importance in the recent observational context, as most of the wave parameters are in accordance with what PSP and SO measured during their near-Sun data acquisition. 
Various observational studies prove that a very high percentage of the whistler waves measured in the young solar wind propagate in a quasi-parallel direction to the magnetic field \citep[e.g.][]{Jagarlamudi2021, Froment2023, Choi2024} and that a large proportion of them propagate in the anti-sunward direction \citep{Kretzschmar2021, Colomban2024}. 
\cite{bervcivc2021whistler}, analysing SO data, find a direct correlation between the detection of whistler waves and the presence of the deficit in the electron VDF. Notably, the waves observed by \cite{bervcivc2021whistler} are also predominantly quasi-parallel and propagate away from the Sun. 
The waves generated during the instability proposed and studied in this work have proprieties matching those of the recent observational studies \citep[e.g.][]{Froment2023, Colomban2024, Choi2024}, but also of many other less recent observations \citep[e.g.][]{Lacombe2014, Stansby2016, Tong2019}.

We also show how the whistler waves, propagating away from the Sun, resonate predominantly with sunward electrons (travelling in the opposite direction to the waves). This ensures that these waves do not interact with the electron strahl. However, wave-particle interactions lead to the filling of the deficit. As the initial deviation from thermodynamic equilibrium is reduced, a decrease in the electron heat flux (defined as the third moment of the VDF) occurs.
According to \citealp{Halekas2020b}, while the drifting electron core represents the sunward contribution of heat flux in the solar wind, the deficit and strahl represent the generally larger anti-sunward contribution. So, although anti-sunward parallel whistler waves do not interact with the strahl, they are able to suppress part of the heat flux that the solar wind carries. The instability described in this letter adds to the possible non-collisional mechanisms responsible for heat flux regulation, as also described by \cite{Coburn2024} in their analysis of SO measurements. This is especially valid in the slow solar wind, where the non-thermality of the electron distribution function is not predominantly related to the presence of the strahl as it is in the fast solar wind \citep{Marsch_2004}.

We describe a possible scenario that correlates the different suprathermal species characteristics of the electron VDF, i.e. strahl, halo and deficit, with parallel and oblique whistler waves. We envision a multistep process that can be broken into the following stages:
\begin{enumerate}
\item the strahl generates sunward-directed oblique and parallel whistler waves due to the whistler heat flux instability.
\item Sunward-directed whistler waves scatter the strahl into the halo. During this process, the oblique whistler waves shift towards smaller propagation angles until they become quasi-aligned with the ambient magnetic field. This mechanism of quasi-parallel sunward whistler wave generation is described in detail in \cite{Micera2020b, Micera_2021}.
\item Anti-sunward quasi-parallel whistler waves are triggered by the interplay of strahl and halo with a further feature of the electron VDF, the deficit. While the strahl population is scattering in the halo, there is a resulting configuration, reproduced here, comprised of strahl-halo-deficit that leads to the instability analysed in this letter.
\item The deficit is filled as a result of the resonant interaction between the electrons adjacent to it and the anti-sunward quasi-parallel whistler waves generated by the instability.
\end{enumerate}

We have thus provided an overall picture of the types of processes that can produce the range of whistler waves observed in the solar wind, without having to call into question the whistler anisotropy instability \citep{Vasko2019} or the electron firehose instability.
The whistler anisotropy instability is an unlikely candidate for whistler wave production in the solar wind because it requires values of $T_{e, \perp} > T_{e, \parallel}$ to be triggered (rarely measured in conjunction with the observation of whistler waves \citep[e.g.][]{Stansby2016}). \\
The electron firehose instability, instead, generates low-frequency left-hand polarised waves \citep{Micera2020} or non-propagating waves ($\omega_r = 0$) \citep{Camporeale2008, Lopez2022}.

In validating and comparing our findings against current observations, three recent studies are noteworthy.
\cite{Cattell2022} discuss the lack of clear evidence of whistler waves when the PSP samples the solar wind inside approximately $25~ R_{s}$.\\
\cite{Halekas_2021_deficit} show that while the deficit is an almost ubiquitous feature of the pristine solar wind in which the PSP is immersed, the clear evidence of this feature fades as the spacecraft moves farther from the Sun.\\
\cite{Choi2024} observe that most of the whistler waves in the young solar wind between $25 - 40 ~ R_{s}$ propagate toward the Sun, while an increasing occurrence of anti-sunward propagating whistler waves is observed between $40 - 55 ~ R_{s}$. Furthermore, consistent with \cite{Cattell2022}, a sharp decrease in the occurrence of whistler waves is noted around $25 ~ R_{s}$.\\
These observations can be explained as follows:
in the inner heliosphere, the electric potential is large, hence the deficit is commonly present. The suprathermal population consists mostly of the strahl, as the relative density of the halo near the Sun is negligible \citep{Halekas2020}. The strahl is not yet unstable due to the low value of its drift velocity compared to the Alfvén velocity \citep{Lopez2020, Micera_2021}. 
Hence, the production of whistler waves by the strahl is suppressed.\\
Additionally, no "scattered electrons" are present in the phase space region characterized by $v_{\parallel} < 0$. In this configuration, all whistler waves interacting with the deficit at $v_{\parallel} < 0$ undergo damping and are, therefore, not observable. It is only in the presence of scattered electrons with $v_{\parallel} < 0$ that whistler waves can grow and become detectable. Thus, in regions where only the "core" component exists, and no "scattered electrons" are present at $v_{\parallel} < 0$, anti-sunward parallel whistler waves are significantly damped due to cyclotron resonance with core electrons.\\
This provides an explanation for the diminishing whistler wave activity closer to the Sun, consistent with the observations reported in \cite{Cattell2022} and \cite{Choi2024}.\\
As the distance from the Sun increases, the Alfvén velocity decreases, making the strahl unstable to the oblique whistler heat flux instability \citep{Verscharen_2019, Micera_2021}. This instability leads to the generation of predominantly sunward-propagating whistler waves.\\
Further away from the Sun, the scattering processes between the strahl and sunward whistler waves become more significant, resulting in electron distribution functions similar to those simulated in this study. 
These distributions can be considered a primary source of anti-sunward whistler waves. This entire process leads to an increase in the relative density of the halo at the expense of the strahl and to the filling of the deficit as the heliocentric distance increases, in agreement with \cite{Halekas_2021_deficit}.

The linear analysis of this new instability would help to understand the physics of the shown instability more clearly. Due to the very non-Maxwellian shape of the distribution, classical solvers to the linear Vlasov-Maxwell dispersion relation cannot evaluate the linear stability of this system. The code ALPS \citep{Verscharen_ALPS}, which specifically addresses the stability of system characterized by non-Maxwellian distribution functions, will be used for a follow up analysis.

To conclude, this work aims at addressing one of the most fascinating topics in heliospheric physics: the link between global and kinetic scales. The electron deficit is a consequence of global electron circulation patterns, while the instability that erases it is a microscope kinetic instability linked to the process of strahl-to-halo scattering, which is in turn associated with a process of global significance such as heat flux regulation in the heliosphere \citep{Verscharen2019_book}.

\acknowledgments

A.M. is supported by the Deutsche Forschungsgemeinschaft (German Science Foundation; DFG) project 497938371.
M.E.I. acknowledges support from DFG within the Collaborative Research Center SFB1491.
D.V. is supported by STFC Consolidated Grant ST/W001004/1. This research was supported by the International Space Science Institute (ISSI) in Bern, through ISSI International Team project 529 (Heliospheric Energy Budget: From Kinetic Scales to Global Solar Wind Dynamics) led by M.~E.~Innocenti and A.~Tenerani.
The authors gratefully acknowledge the Gauss Centre for Supercomputing e.V. (\url{www.gauss-centre.eu}) for funding this project by providing computing time on the GCS Supercomputer SUPERMUC-NG at Leibniz Supercomputing Centre (\url{www.lrz.de}).

\bibliographystyle{aasjournal}  
\bibliography{main}

%\listofchanges
\end{document}